# Topics and Treatments in Global Software Engineering Research
# A Systematic Snapshot

Bilal Raza, Stephen G. MacDonell and Tony Clear
SERL, School of Computing & Mathematical Sciences, Auckland University of Technology,
Private Bag 92006, Auckland 1142, New Zealand

**Abstract**

*This study presents an analysis of the most recent literature addressing global software engineering (GSE). The primary purpose is to understand what issues are being addressed and how research is being carried out in GSE – and comparatively, what work is not being conducted. We examine the current state of GSE research using a new Systematic Snapshot Mapping (SSM) technique. We analysed 275 papers published between January 2011 and June 2012 in peer-reviewed conferences, journals and workshops. Our results provide a coarse-grained overview of the very recent literature addressing GSE, by classifying studies into predefined categories. We also follow and extend several prior classifications to support our synthesis of the data. Our results reveal that, currently, GSE studies are focused on Management and Infrastructure related factors rather than Human or Distance related factors, using principally evaluative research approaches. Most of the studies are conducted at the organizational level, mainly using methods such as interviews, surveys, field studies and case studies. We use inter-country network analysis to confirm that the USA and India are major players in GSE, with USA-India collaborations being the most frequently studied, followed by USA-China. Specific groups of countries have dominated the reported GSE project locations (and the locations of research authors). In contrast, regions including Central Asia, South Asia (except India), Africa and South East Asia have not been covered in these studies. While a considerable number of GSE-related studies have been published they are currently quite narrowly focused on exploratory research and explanatory theories. The critical research paradigm has been untouched, perhaps due to a lack of criteria and principles for carrying out such research in GSE. An absence of formulative research, experimentation and simulation, and a comparative focus on evaluative approaches, all suggest that existing tools, methods and approaches from related fields are being tested in the GSE context. However, these solutions may not scale to cover GSE-related issues or may overlook factors/facets specific to GSE.*

**Keywords:** Global Software Engineering (GSE), Distributed Software Development, Systematic Mapping.

## 1. INTRODUCTION

Global software engineering (GSE) is a growing field as is clearly evident in the diversity of locations involved and the rapidly increasing number of published studies into GSE-related issues. As the number of such studies increases it becomes important to periodically summarize the work and provide overviews of the results (Petersen et al., 2008), as a means of reflection on what work is being done and what gaps might exist. Various fields have specific methodologies to carry out such secondary studies (Petersen et al., 2008). Often in these studies, at least in the software engineering (SE) domain, the evidence pertaining to a specific topic or research question is investigated over a period of five to ten years. In this paper, we utilize a different approach - we investigate the breadth of topics covered over a short timeframe, an approach we refer to as a systematic snapshot. This establishes a baseline state that could be extended in a backward or forward direction to analyse changes over time. The systematic mapping (SM) method has been widely used in medical research (Petersen et al., 2008) and was first adopted in software engineering research by Bailey et al. (2007). SM aims to provide high-level analysis of relevant research literature by classifying the work according to a series of defined categories and visualizing the status of a particular field of research (Mohd Fauzi et al., 2010, Petersen et al., 2008). This technique has been used recently in the GSE field (Steinmacher et al., 2012, Jalali and Wohlin, 2010, da Silva et al., 2011, Mohd Fauzi et al., 2010, Portillo-Rodríguez et al., 2012). In these studies specific aspects of GSE research were categorized (using guidelines presented in (Kitchenham and Charters, 2007, Petersen et al., 2008)). These studies considered between 24 and 91 papers (24, 91, 77, 70 and 66 studies, respectively) up to the year 2010 in their final analyses. The aspects of GSE analysed in these studies were software configuration management, awareness support, agile practices, project management, and tools in GSE. All five studies therefore classified the GSE literature from a relatively narrow perspective but covering a wide temporal range. They were published in well-known journals and conferences and provide valuable contributions to the body of GSE literature. In our study, we instead use a new systematic mapping process called Systematic Snapshot Mapping (SSM), briefly described in section 3, to classify the very current global software engineering literature.

The next section provides a brief background to related studies, and section III describes our method. In the subsequent section IV our results are presented followed by a discussion of validity threats in section V. In section VI we conclude this paper and section VII conveys future work.

## 2. BACKGROUND AND RELATED MAPPING STUDIES

Interest in software development carried out by globally distributed, culturally and/or temporally diverse teams arose with the advent of outsourcing in the last two decades and it continues to increase (Šmite et al., 2010). Its importance has led to the specific area of research and practice called global software engineering (GSE) (Šmite et al., 2010). In a recent review Šmite et al. classified the empirical GSE research, considering studies published between 2000 and 2008, and presented the results in two papers (Šmite et al., 2008, Šmite et al., 2010). They concluded that GSE was (still) an immature field with limited empirical studies. They further concluded that the majority of studies focused on different aspects of GSE management rather than the in depth analysis of GSE solutions.

Jalali and Wohlin (2010) reported a SM study based on their analysis of 77 studies published between 1999 and 2009. They focused their work on the application of agile practices in GSE and explored under which circumstances these practices have been used successfully in that context. The results reveal that in most cases agile practices were modified based upon the context and requirements. The authors also expressed the need for integrating experiences and practices to assist practitioners. da Silva et al. (2011) presented an evidence-based project management model for distributed software development based on the synthesis of 70 papers published between 1997 and 2009. They aimed to provide feedback to help practitioners and researchers understand challenges and implement effective solutions to improve project management in distributed settings. Fauzi et al. (2010) presented the results of a SM study of software configuration management (SCM) in GSE. They found that a lack of group awareness and coordination exacerbates the issues of SCM and no process had been proposed to address this. Their review considered 24 papers published between 1999 and 2010. Rodriguez et al. (2012) conducted a SM study, analysing 66 papers published between 2000 and 2010. They compiled a list of 132 tools used in global software projects and classified them to help practitioners and researchers make use of the available tool support. It was found that the majority of these tools had been developed at research centres and just 19% were reported to have been tested outside the context in which they were developed. Another SM study was reported by Steinmacher et al. (2012). In this paper they reviewed 91 studies regarding awareness support in distributed software development (DSD). They found that coordination is the most supported dimension of the 3C model whereas communication and cooperation are less frequently explored. All of the above mentioned SM studies provide valuable contributions to the body of GSE literature and include content intended to support practitioners. Each addresses a specific aspect of GSE and considers around a decade of research in the field. In our study we covered a shorter time period using a different approach, described in the next section.

## 3. METHOD AND CONDUCT

The results presented in this paper correspond to our classification of the current literature on GSE. We used a new method for carrying out SM studies called Systematic Snapshot Mapping (SSM). In order to classify the current literature, we chose the time period between January 2011 and June 2012. This study followed guidelines presented by Petersen et al. (2008) for carrying out systematic mapping studies. However, instead of narrowing down the topic and considering a large temporal period, we limited the time span and considered the full breadth of topics covered. This study was inspired by several prior classifications of SE and GSE literature including that of Glass et al. (2002), but instead of following a random sampling technique to select papers (as in (Glass et al., 2002)) we used a systematic process. We followed the general guidelines of (Petersen et al., 2008) and employed a defined protocol for choosing search strings and executing them against relevant databases to cover the breadth of GSE-related studies. Thus, instead of limiting the topic itself (as per the SMs cited above) we limited the scope by using a small temporal range, giving us an up-to-date snapshot overview of the literature. We defined our categories at the outset of our analysis and chose various dimensions to present the results, mainly leveraging the prior classification work of Richardson et al. (2012) and Glass et al. (2002). We present our results in the form of tables, bar graphs, bubble plots and network analysis graphs to provide visual representations of the data. We believe such a snapshot approach is especially useful in cases where a field is moving rapidly and where there is consequently rapid growth in the research literature. This new approach for carrying out SM also provides an opportunity to effectively build upon different researchers' work by using different temporal ranges. Since traditional approaches such as systematic literature reviews and systematic mappings use narrowly defined topics it is difficult to analyse how overall trends evolve over a period of time. This study provides a baseline against which analyses using other temporal ranges could be compared.

### 3.1 Research Questions
In order to present a current snapshot of the GSE research literature, the following research questions were established for this study:

**RQ1.** What are the factors, levels and locations investigated in the current GSE literature?

**RQ2.** How is the current research being carried out in GSE in regard to methods and approaches?

## 3.2 Search Strategy

Our search strategy was designed to keep the topic general while addressing a short time period to provide an up-to-date overview of the research literature. Initial search keywords were selected from known GSE systematic literature reviews and mapping studies. These keywords were updated based upon various dry runs carried out on the Scopus database to ensure their effectiveness. In the initial run, a target was set to ensure at least those studies from which the keywords were taken were retrieved. In the second run, a random set of ten studies was selected from the Proceedings of the 2009-2011 ICGSE conferences, and the search strings were further refined to ensure that these sample studies were also retrieved. A similar method was used by Jalali and Wohlin (2010) to justify and improve the utility of their selected key terms.

Table 1: List of keywords used as search strings.

```
"global software engineering" OR "global software development" OR
"distributed software engineering" OR "distributed software
development" OR "offshore software development" OR "offshore
software engineering" OR "distributed team" OR "global team"
"offshore insourcing" OR "geographically distributed teams" OR
"global software" OR {("software" OR "information system" OR
"computer" OR "information technology") AND ("virtual team" OR
"dispersed team" OR "far shore" OR "offsite" OR "offshore
outsource" OR "outsourced"}|
```

Table 1 shows the final list of keywords used to cover as many variations of the same term as possible. As this area of research is still maturing, we intentionally adopted many keywords having low precision but high recall (Dieste and Padua, 2007).

## 3.3 Data Sources

We searched across multiple data sources to retrieve as many potentially relevant studies as possible. Initial preference was given to the use of the electronic database Scopus as it provides comprehensive coverage of relevant GSE journals. It is especially recommended for the software engineering and computer science fields as it covers many of the well-known publishers in these disciplines. Simultaneously, IEEE Xplore, the ACM Digital Library, SpringerLink and ScienceDirect were also searched to complement the Scopus results. Each database has its limitations in terms of the number of keywords accepted at a specific instance; therefore, we had to break the search phrases to suit the particular database. These subsequent searches, which tended to find limited additional studies to those found via Scopus, added to our confidence that relevant studies had not been missed in our search.

## 3.4 Data Retrieval

Data was retrieved in multiple steps. In the first step, citations of retrieved studies were downloaded and stored as separate EndNote files based upon their abstracts and titles. After this process, all the EndNote files were combined and duplicate papers were removed. Once all the duplicates were removed the studies were then considered for the inclusion process. The search and retrieval process was conducted in July 2012 and the date range was limited to January 2011 to June 2012. The search was carried out on metadata (title, abstract, keywords) and only peer-reviewed literature published in English was considered.

## 3.5 Inclusion Process

The steps taken in the inclusion process to select studies are shown in Figure 1. After searching each database 2020 studies were retrieved. The decision for further analysis of studies was based upon the first author's reading of the papers' titles or abstracts (resulting in 1125 studies). After this step, duplicates were removed using the 'Find Duplicates' feature of EndNote – this led to the removal of more than half of the studies under consideration. After removing the duplicates full text versions of each study were sought. For 12% of the papers (53 of the 437 remaining) the full text was not available to us, primarily because the papers were not published in well-known journals or conference proceedings. These studies were not considered for further analysis. The full text of the remaining 384 papers was then reviewed by the first author and a final set of 275 studies was selected for inclusion in the SM analysis. In this stage, studies in the form of short papers, extended abstracts and position papers (only describing future work) were excluded. A number of studies, not related to the software engineering domain, had slipped through to this stage and upon cursory review of the full text were also excluded.

## 3.6 Data Extraction and Synthesis

We followed the general guidelines provided in (Petersen et al., 2008) to build a classification scheme. The included studies were categorized according to various dimensions: research approach, research method, GSE factors, level of analysis and GSE locations. In order to reduce the threats to validity, regular meetings of the three authors were held to discuss issues and address misconceptions. In order to reduce bias effects the three researchers also conducted a sample classification together. At a later point a further sample of studies which were initially classified by the first author were verified by the senior researchers, discussions were held again and issues were addressed. It was established that the authors were in general agreement regarding the classification, based upon the sample results.

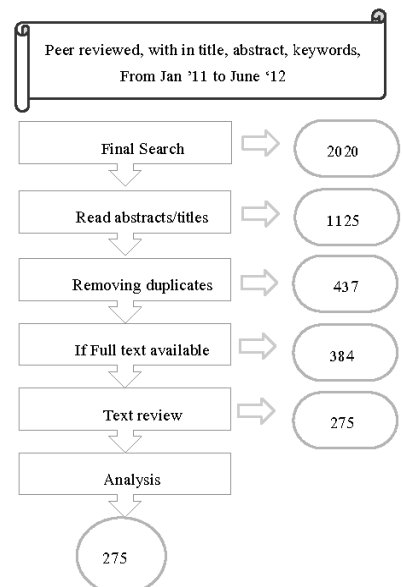

Figure 1: Inclusion process.

The classification scheme utilized by Glass et al. (2002) was used to characterize the research approach for our set of studies. This scheme (Glass et al., 2002) was based on an earlier categorization by Morrison and George (1995) in which the following four approaches were employed: Formulative, Evaluative, Descriptive and Developmental. The first three categories were further divided to provide a rich set of candidate research approaches (Glass et al., 2002). The descriptive and formulative categories characterize non-empirical studies (Glass et al., 2002). The descriptive category has three subcategories. Subcategory Descriptive-System was used to capture papers focused on describing a system, Descriptive- Other was used to categorize papers that included an opinion piece, and the third descriptive category used was Review of literature (Glass et al., 2002). The formulative approaches were classified into six subcategories to cover the major entities being formulated: Framework, Guidelines/standards, Model, Process/method, Classification/taxonomy and Concept (Glass et al., 2002). The Evaluative category in (Glass et al., 2002) drew on three main research epistemologies identified by Orlikowski and Baroudi in (1991): Positivist as Evaluative- Deductive, Interpretive as Evaluative-Interpretive, and Critical as Evaluative-Critical. Evaluative-Other (Glass et al., 2002) was added in this list to include studies that have an evaluative component but did not use any of the above approaches e.g. opinion surveys. So in total, we used these 13 subcategories formulated in (Glass et al., 2002) to classify the research approaches used in our set of GSE studies. It was found that epistemologies were rarely mentioned in the Abstract or Introduction of the studies and various other sections had to be traversed to enable this particular classification. Glass et al. (2002) also encountered this same issue.

We also mainly considered (Glass et al., 2002) for the list of methodologies used in software engineering research. However, to better reflect the GSE perspective we also considered the methodologies considered in (Šmite et al., 2008, Šmite et al., 2010). We added computer mediated communication (CMC) analysis to cover studies that investigate artefacts such as chat-histories and emails. Although grouped together in prior studies, Observations and Interviews were considered separately, as many studies use them to complement other methods. Interviews are widely used as a sub- method in Case Studies and Observations are used in Ethnographies. However, we observed that these methods are being used in their own right and we therefore classified them separately. We included the method Data Analysis to signify studies that utilized data from Repositories, Incident Management Systems and Archives of previous projects. We used Proof of Concept for non-empirical studies in which entities were formulated but were only described by examples rather than any formal validation. Some other generic data along with the above categorizations were captured in a spread sheet:

- Title of the paper
- Author (s) and their geographic location
- Summary of the conclusion
- The source (journal, conference or workshop)
- Process activities/artefacts, practices (if discussed)
- Type of contribution
- Geographical locations of mentioned projects.

# 4. FINDINGS

This section presents the results obtained based on the data extracted from our final set of 275 studies.

## 4.1 Findings for Factors

Richardson et al. (2012) identified 25 GSE factors in an empirical study and grouped them in the four broad categories of Distance, Infrastructure, Management and Human Factors. We used these categories to also characterize our identified studies. We added Learning/Training/Teaching, Competition and Performance to the Management category and Relationship to the Human Factors category. We also updated the latter category with Coordination/collaboration. Table 2 presents the results of this classification. The results clearly show that current GSE studies are heavily focused on Management and Infrastructure related factors compared to Human and Distance related factors. Šmite et al. in (2010) presented a systematic review of empirical GSE studies and also found that most of the studies were focused on management related issues. Comparing these results with the SWEBOK (Alain et al., 2001) knowledge areas (KAs), it was found that the standard lacks specific considerations for GSE. As a corollary, it was also found that KAs related to design, construction, testing and maintenance are not widely addressed in the recent GSE literature.

Table 2: Findings for GSE factors.

| GSE Factors | Percentage |
|---|---|
| *Distance* | *17.5%* |
| Communication | 8.9% |
| Language | 1.1% |
| Culture | 5.5% |
| Temporal issues | 1.8% |
| *Human Factors* | *14.7%* |
| Fear | 0.5% |
| Motivation | 2.3% |
| Trust | 2.7% |
| Cooperation | 1.8% |
| Coordination/collaboration | 5.9% |
| Relationship | 1.4% |
| *Management* | *44.5%* |
| True Cost | 1.8% |
| Project Management | 8.9% |
| Risk Management | 2.3% |
| Defined roles and responsibilities | 1.6% |
| Team Selection | 0.9% |
| Effective Partitioning | 4.6% |
| Skills Management | 0.4% |
| Knowledge transfer/knowledge | 6.7% |
| Visibility | 3.4% |
| Reporting Requirement | 0.0% |
| Information Management | 1.1% |
| Teamness | 5.5% |
| Learning/Training/teaching | 4.6% |
| Competition | 0.6% |
| Performance | 1.6% |
| *Infrastructure* | *23.3%* |
| Process Management | 7.1% |
| Tools | 8.5% |
| Technical Support | 0.4% |
| Communication tools | 7.1% |

## 4.2 Findings for Research Approach

GSE presents a complex context that demands a more extensive repertoire of research methods and approaches than those currently prevailing (Clear and MacDonell, 2011). Table 3 presents the findings of the classification of research approaches used in current GSE-related studies. In terms of the three main categories, the dominant research approach is Evaluative, followed by Descriptive and then Formulative. This is in sharp contrast to the results reported in 2002 by Glass et al. in which the order was Formulative, Descriptive and Evaluative. One of the main reasons for the present dominance of Evaluative research is the inclusion of new empirical methods such as CMC analysis, Interviews, Data Analysis and Observations. These results appear to be in contrast with the results of Šmite et al.'s systematic review (2010) of GSE-related studies published between 2000 and 2008. They concluded that GSE-related studies are relatively small in number and immature and most of them focused on problem-oriented reports. Our current results show, however, that GSE publications have grown in quantity and quality and more studies have used evaluative approaches. Of note is these evaluative approaches are mostly confined to previously formulated work. We interpret this to mean that existing methods, tools and so on from related fields, such as collocated software engineering (CSE), are being evaluated in the context of GSE.

Given that GSE is fundamentally different from CSE (Richardson et al., 2012), it seems likely that solutions formulated for CSE will need to be updated or enhanced for GSE. Entirely new solutions may also need to be identified and assessed in the GSE context. Similarly, there is clear potential for critical research in this context particularly in light of the power structures that can exist between GSE 'partners', and the associated issues of trust, fear, cooperation and the like (as shown in Table 2). Criteria or principles for carrying out critical research are lacking generally in information systems (IS) (Myers and Klein, 2011). Considering its importance, Myers and Klein in (2011) proposed a set of principles for conducting critical research - these principles could be considered in future investigations of human factors in GSE.

## 4.3 Findings for Research Methods

Figure 2 depicts the research methods used. The most dominant methods are Interview, Survey, Field Study and Case Study, indicating that most of the studies employed qualitative methods. These results are also in stark contrast to (Glass et al., 2002) in which SE researchers used very few case or field studies. For studies in which multiple methods were used we assigned more than one research approach and method. Research methods in GSE are currently skewed towards exploratory research focusing on theories relating to 'Explanation' as described by Gregor (2006). These theories aim to provide explanation about what, how and why things happen and to promote greater understanding of phenomena. Thus, although GSE research has grown in terms of the number of studies being conducted, these studies are exploratory and/or explanatory in nature. It will be interesting to compare these results with future studies to determine whether work moves towards more predictive studies as the field matures.

Table 3: Findings for research approach.

| Research Approach | Percentage |
|---|---|
| *Descriptive* | *25.4%* |
| Descriptive-system (DS) | 7.8% |
| Review of literature (DR) | 9.7% |
| Descriptive-other (DO) | 7.8% |
| *Evaluative* | *56.2%* |
| Evaluative-deductive (ED) | 17.1% |
| Evaluative-interpretive (EI) | 26.4% |
| Evaluative-critical (EC) | 0.0% |
| Evaluative-other (EO) | 12.7% |
| *Formulative* | *18.3%* |
| Formulative-framework (FF) | 5.1% |
| Formulative-guidelines/standards/approach | 1.9% |
| Formulative-model (FM) | 5.6% |
| Formulative-process, method, algorithm (FP) | 3.1% |
| Formulative-classification/taxonomy (FT) | 0.7% |
| Formulative-concept (FC) | 1.7% |

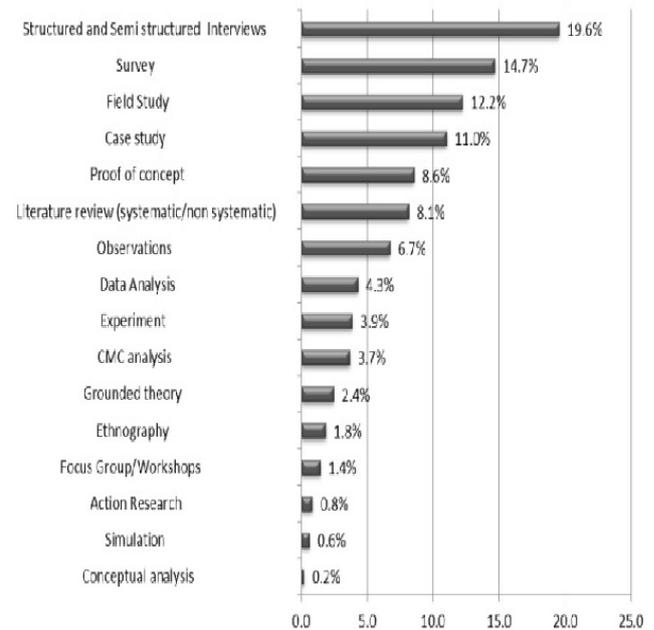

Figure 2: Findings for research methods.

## 4.4 Findings for Level of Analysis

Figure 3 shows the level of analysis considered currently by GSE researchers. The dominant level of analysis was found to be Organizational followed by Inter-Organizational - combined together they are used in more than half the studies reviewed. Fewer studies addressed group, individual and societal levels, a finding that coincides with the results of Glass et al. (2002) in respect of SE studies.

## 4.5 Findings for Distribution of Studies

Table 4 presents the distribution of studies across various conferences, journals and workshops with frequency greater than one. (This limit was imposed due to space considerations and for ease of interpretation.) The majority of the selected studies were published in conference proceedings and drew on an industrial context.

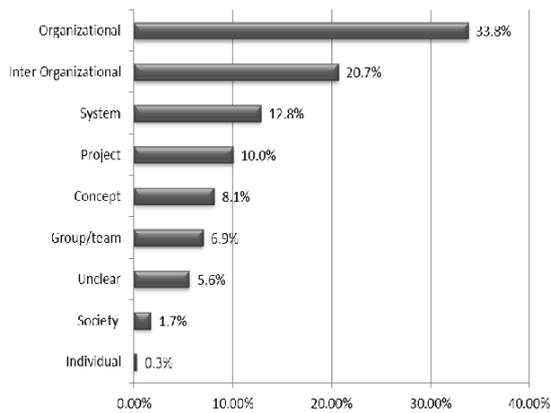

Figure 3: Findings for level of analysis.

## 4.6 Bubble Plot Analysis

The use of visual techniques, such as bubble plots, has been recommended by (Petersen et al., 2008) and such techniques have been used to convey the results of mapping and classification studies (Šmite et al., 2008, Jalali and Wohlin, 2010). Figure 4 presents the results of this study in the form of a bubble plot. We chose to represent three classifications within it: Research approach is on the right X-axis, GSEfactors, grouped in their four major categories, are on the Y-axis, and level of analysis is on the left X-axis. The results clearly show that most of the current studies are focused on using evaluative approaches around management and infrastructure factors and analysed at the organizational levels. Studies based upon specific groups, societies and individuals are found to be limited. Organizational concerns have been at the forefront in terms of the level of analysis, leaving much scope for consideration of groups and individuals for future studies.

Table 4: Distribution of studies across Journals, Conferences and Workshops.

| Journals | | CSCW | 8 |
|---|---|---|---|
| IST Journal | 8 | PROFES | 6 |
| JSEP | 7 | CHI | 5 |
| J SOFTW MAINT EV | 7 | XP | 4 |
| IET Software | 6 | ICIC | 3 |
| J of E MARKETS | 4 | PICMET | 3 |
| IEEE Software | 4 | ISEC | 3 |
| J COMM and COM SC | 3 | ICSSP | 3 |
| ISJ | 3 | MySEC | 2 |
| IJoPM | 2 | EUROMICRO | 2 |
| JSW | 2 | ICIS | 2 |
| POM Journal | 2 | CollaborateCom | 2 |
| IS | 2 | CTS | 2 |
| IEEE TEM | 2 | PACIS | 2 |
| LNBIP | 2 | **Workshops** | |
| J Grp Dec Negot | 2 | CTGDSD | 13 |
| **Conferences** | | ICGSE | 13 |
| ICGSE | 26 | CHASE | 7 |
| HICSS | 15 | OTM | 3 |
| ICSE | 8 | Global Sourcing | 3 |

## 4.7 Location of GSE Projects

Figure 6 and Table 5 provide graphical and tabular representations of the locations involved, whereas Figure 5 depicts the spread of the number of locations involved in GSE projects. Figure 5 shows that most of the studies are focused on projects involving two locations. A few studies also mentioned regions rather than countries; we considered them in the next section.

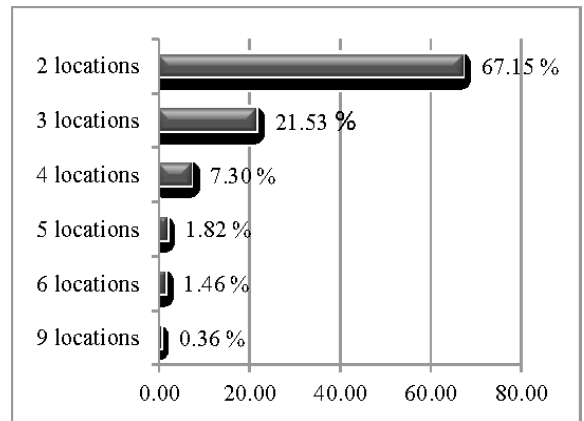

Figure 5: Number of locations used in GSE projects.

In Figure 6, countries and regions marked by darker shades arethose most frequently involved in GSE. For ease of analysis we grouped these countries into six categories based upon the number of studies that cite their involvement in global projects. Not unexpectedly, the two countries reported as most frequently involved in global software projects are the USA and India. Countries including Germany, Finland, China, the UK, Australia and Brazil are ranked in the second group, closely followed by a group comprising Sweden, Japan, Argentina, the Netherlands, Spain, Canada and Switzerland. In the next two categories lie the potentially upcoming and emerging countries of Russia, Eastern European countries such as Lithuania, Far Eastern countries including Malaysia and Indonesia, and the South/Central American countries of Chile and Mexico. These representations give some insight into the diversity of countries' involvement in GSE projects. Some of these regions are underrepresented but this does not necessarily mean that these locations are not involved in GSE; it could be that these regions have simply not been considered in recent studies. Researchers often seek industrial contacts to validate their research outcomes and gather feedback to improve their results, and often times they rely on personal contacts in their national industries. Our study also shows that the top seven locations of GSE authors are the USA, Finland, Germany, Spain, Brazil, India and Sweden. Apart from Spain, which is thirteenth, all six other countries are in the list of top ten locations involved in GSE projects.

## 4.8 Inter-country Relationship Analysis

Figure 7 shows the results of our inter-country network analysis. We used NodeXL, an extendable tool kit used for data analysis and visualizations andan add-in to Microsoft Excel spread sheet software (Smith et al., 2009), to support our analysis. Table 6 lists the pairwise relationships with frequency greater than one. (This constraint was imposed due to space limitations; however, all the relationships are shown in Figure 7.) It can be seen in Figure 7 and Table 6 that the most connected nodes are the USA and India. Some studies explicitly mentioned the collaborating locations whereas others only specified the locations involved without clearly stating which actively collaborated. For the latter studies, we assumed pairwise relationships between each location. For future studies we recommend that authors clearly state the nature of each party's involvement. A few studies also mentioned regions rather than countries; we used these as needed in Figure 7.

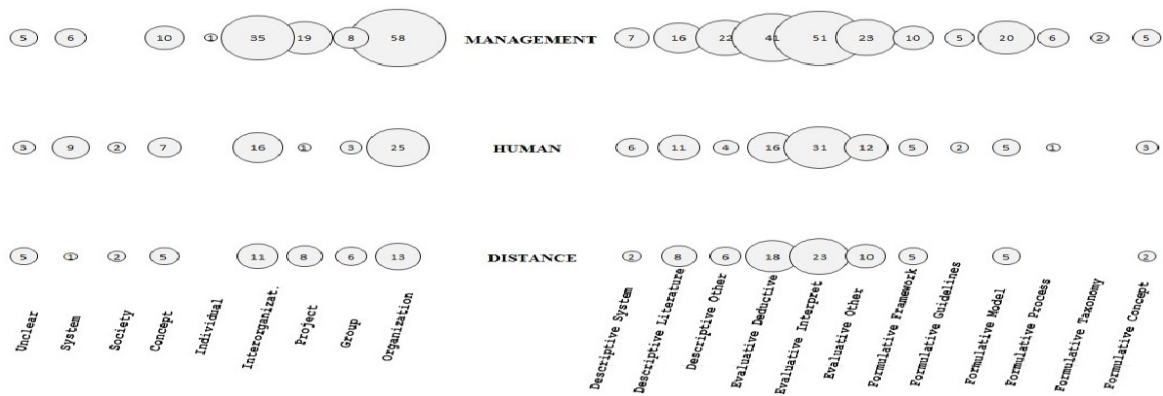

Figure 4: Bubble plot analysis.

## 5. THREATS TO VALIDITY

One of the main threats to the validity of our study is the incomplete selection of primary studies or missing relevant studies. There is a possibility that, even though we followed a systematic process, we may have missed some related work. In order to mitigate this risk we formulated a wide variety of search-terms. These terms were taken from related SM/SLR studies and were updated based upon the retrieved results. Initially, we ensured that at least those SM/SLR studies were indeed retrieved using the search terms drawn from each study. In the next stage, we constructed a sample list of studies from various ICGSE proceedings and ensured that the search terms retrieved these studies as well. During this process the search terms were continuously updated until all sample studies were retrieved, similar to the approach taken by (Jalali and Wohlin, 2010). A second validity threat arises due to researcher bias during the classification process. In order to reduce this threat, we carried out some sample classifications collectively. Furthermore, the lists of studies as classified by the first author were validated by the senior researchers involved. A high level of agreement was achieved, giving us confidence that the classification process was executed appropriately and consistently.

## 6. CONCLUSIONS

Through this study we have provided a current snapshot of the GSE-related literature. We classified 275 empirical and non-empirical studies, published between January 2011 and June 2012, into predefined categories (see http://tinyurl.com/GSE-Papers). We examined the following characteristics: GSE factors, research approaches, research methods, level of analysis, and GSE project locations. The GSE factors most frequently researched were related to management and infrastructure using evaluative approaches and taking an organizational perspective as the level of analysis. Regarding research methods, interviews, surveys, case studies and field studies are the most commonly used. In relation to project locations, the USA and India are the predominant nations involved in global software projects. Inter- country network analysis also shows that USA-India collaboration is at the top followed by USA and China. It will be interesting to carry out further similar snapshot studies on an on-going basis to see if or how these trends evolve. Similarly, studies could be carried out retrospectively on previous years' research literature to enable comparisons with this study. This study aims to provide a stepping stone for further related studies.

Table 5: Locations involved in GSE projects.

| Country | Frequency | Country | Frequency | Country | Frequency |
|---|---|---|---|---|---|
| US | 238 | Italy | 11 | Estonia | 4 |
| India | 159 | Norway | 11 | Philippines | 4 |
| Germany | 57 | Czech | 10 | Thailand | 4 |
| Finland | 54 | Lithuania | 10 | Vietnam | 4 |
| China | 44 | Israel | 9 | Korea | 4 |
| UK | 38 | Malaysia | 9 | Costa Rica | 3 |
| Australia | 32 | Mexico | 9 | Colombia | 3 |
| Brazil | 32 | Senegal | 9 | Ecuador | 3 |
| Sweden | 27 | Singapore | 9 | Egypt | 3 |
| Japan | 20 | New Zealand | 8 | Greece | 3 |
| Argentina | 19 | Cambodia | 7 | Poland | 3 |
| Netherlands | 19 | France | 7 | Taiwan | 3 |
| Spain | 18 | Belgium | 6 | Romania | 2 |
| Canada | 16 | Chile | 6 | Slovakia | 2 |
| Switzerland | 16 | Croatia | 6 | Turkey | 2 |
| Ukraine | 15 | Hungary | 6 | Bangladesh | 1 |
| Russia | 13 | UAE | 5 | Pakistan | 1 |
| Denmark | 11 | Panama | 5 | South Africa | 1 |
| Ireland | 11 | Austria | 4 | Tunisia | 1 |

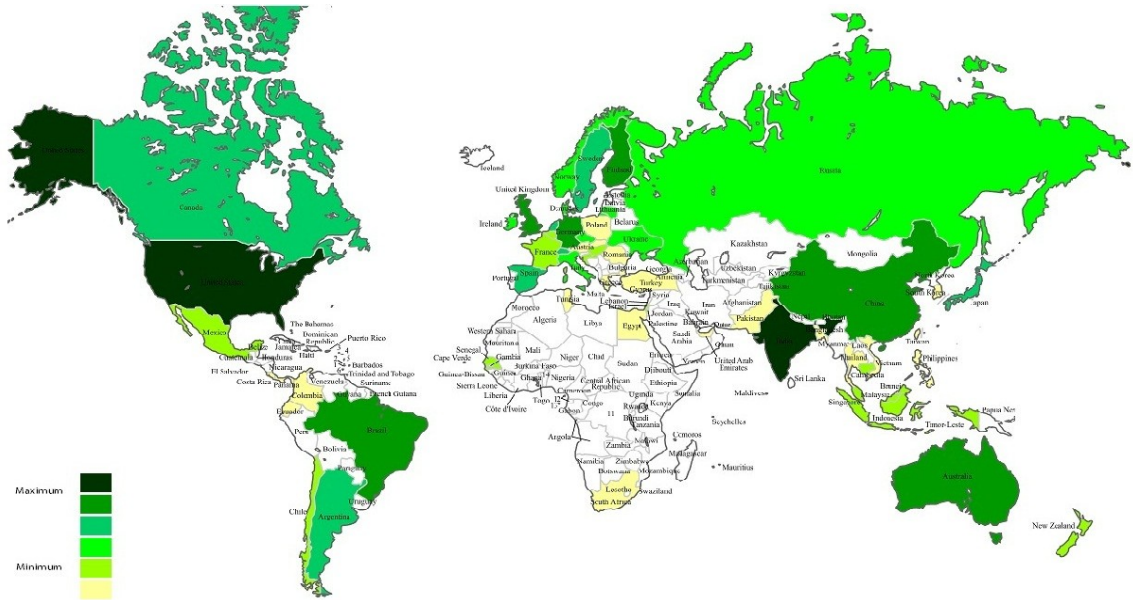

Figure 6: Locations involved in GSE projects.

Table 6: Inter-country relationships.

| | | | | | | | | |
|---|---|---|---|---|---|---|---|---|
| India | US | 67 | Singapore | US | 4 | Australia | Spain | 2 |
| China | US | 23 | UK | Ukraine | 4 | Australia | Germany | 2 |
| Germany | India | 14 | US | Singapore | 4 | Belgium | US | 2 |
| Brazil | US | 11 | US | Russia | 4 | Brazil | UK | 2 |
| Australia | US | 10 | US | Israel | 4 | Cambodia | Senegal | 2 |
| Europe | US | 10 | China | Japan | 3 | Cambodia | India | 2 |
| UK | US | 10 | Denmark | India | 3 | Canada | India | 2 |
| Finland | India | 8 | E. Europe | Finland | 3 | Canada | Europe | 2 |
| Germany | US | 8 | Finland | Sweden | 3 | Europe | Japan | 2 |
| Australia | India | 7 | Finland | Lithuania | 3 | Finland | Japan | 2 |
| India | Europe | 7 | Finland | Baltic C. | 3 | Finland | Brazil | 2 |
| UK | India | 7 | Germany | Russia | 3 | France | Germany | 2 |
| US | Argentina | 7 | Germany | Brazil | 3 | Germany | Czech | 2 |
| Finland | US | 6 | India | Senegal | 3 | India | Switzerland | 2 |
| US | Ukraine | 6 | India | Japan | 3 | India | Middle East | 2 |
| US | Finland | 6 | India | China | 3 | Ireland | China | 2 |
| US | Canada | 6 | Italy | Switzerland | 3 | Lithuania | US | 2 |
| Croatia | Sweden | 5 | Japan | India | 3 | Malaysia | India | 2 |
| Czech | Finland | 5 | Norway | Finland | 3 | Netherlands | Ukraine | 2 |
| Japan | US | 5 | Spain | Germany | 3 | Netherlands | UK | 2 |
| Sweden | Croatia | 5 | US | Switzerland | 3 | New Zealand | US | 2 |
| US | Japan | 5 | US | Sweden | 3 | Norway | Sweden | 2 |
| US | Ireland | 5 | US | Spain | 3 | Norway | Czech | 2 |
| W Europe | India | 5 | US | Senegal | 3 | Spain | Lithuania | 2 |
| Brazil | India | 4 | US | Norway | 3 | Switzerland | Vietnam | 2 |
| Finland | Germany | 4 | US | Mexico | 3 | Switzerland | Ukraine | 2 |
| India | Sweden | 4 | US | Malaysia | 3 | US | Taiwan | 2 |
| India | Argentina | 4 | US | Egypt | 3 | US | Middle East | 2 |
| Netherlands | US | 4 | US | Denmark | 3 | US | Cambodia | 2 |
| Netherlands | India | 4 | Asia | US | 2 | W Europe | US | 2 |

It appears that, in general, existing solutions are being applied in a GSE context, even though these solutions may lack specific considerations needed for GSE. For instance, aspects of non-functional requirements and stage/phase-related issues are not addressed separately in the current GSE literature. Although the field of GSE research has grown rapidly in terms of the number of studies conducted, these studies are quite narrowly focused towards exploratory research and the provision of explanatory theories. Furthermore, in spite of GSE providing a natural and potentially fruitful setting for critical research, such work is yet to be conducted. The current research focus is mainly directed to organizational concerns, leaving much scope for consideration of the needs of stakeholder groups and individuals. The research is also skewed towards projects having two locations showing a dearth of studies relating to multiple locations and their underlying complex relationships. Finally, there are regions of the world that are not being currently studied by researchers and it may be useful to consider them in the future studies, particularly if the dimensions of culture and their impact on GSE are of interest.

## 7. FUTURE WORK

A notable omission in the current focus of work relating to GSE is any sustained coverage of issues to do with power and exploitation. While the human factors tabulated in Table 2 above include some focus on the factors of fear, trust, cooperation and relationship, these are given relatively limited attention. Again in Figure 4 there is a noted absence of studies at an individual unit of analysis. There are no studies giving personal narratives or biographies - are the workers in GSE deliberately kept invisible? Is this absence a function of the research methods used, for instance, no examples of critical evaluative work have been identified in this review? Or is it an abrogation of our duties as academics to act in the role of 'critic and conscience of society'? Will the future see more equal partnerships in sustainable global ventures, or will there be a backlash against crude models of global labour arbitrage? What risks might that pose to a multi-billion dollar industry? These issues warrant more attention by researchers, although difficult to confront. In addition such research will be challenging to design and conduct, yet the absence of critical evaluative studies presents a glaring gap in current GSE research.

## REFERENCES

Alain, A., Pierre, B., Robert, D. & James, W. M. 2001. *Guide to the Software Engineering Body of Knowledge - SWEBOK*, IEEE Press, Piscataway, NJ, 2001.

Bailey, J., Budgen, D., Turner, M., Kitchenham, B., Brereton, P. & Linkman, S. Evidence relating to Object-Oriented software design: A survey. *In: Empirical Software Engineering and Measurement*, 2007. ESEM 2007. *First International Symposium on*, 20-21 Sept. 2007 2007.

Clear, T. & MacDonell, S. G. 2011. Understanding technology use in global virtual teams: Research methodologies and methods. *Inf. Softw. Technol.*, 53, pp.994-1011

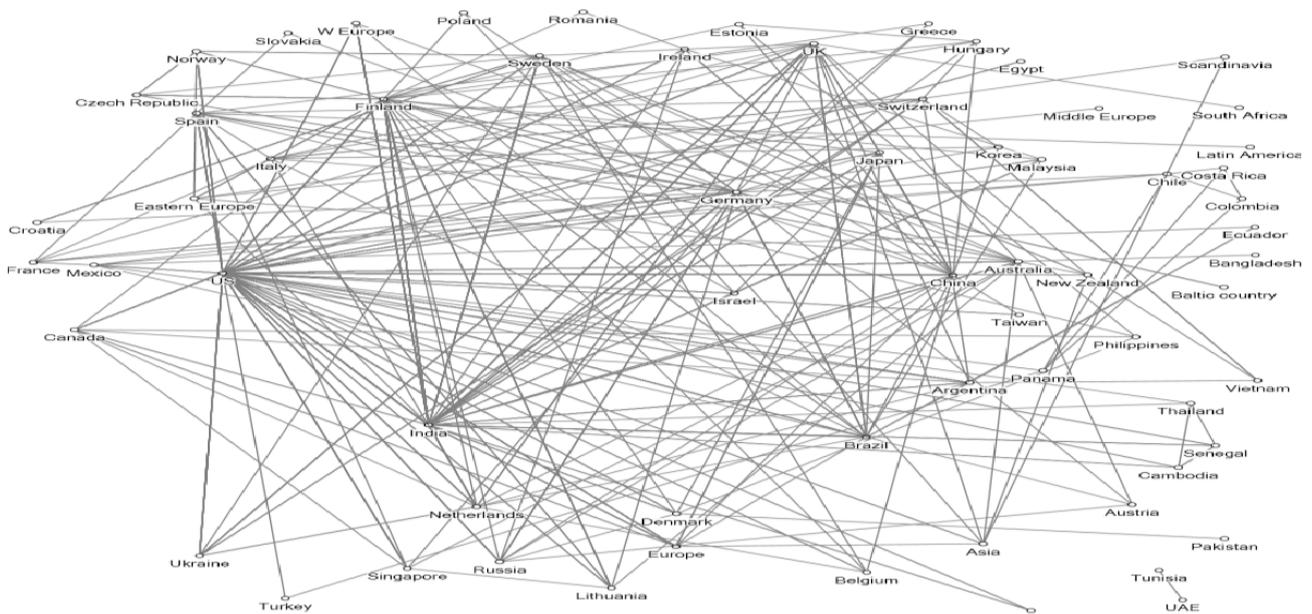

Figure 7: Inter-country relationship analysis.

Da Silva, F. Q. B., Prikladnicki, R., França, A. C. C., Monteiro, C. V. F., Costa, C. & Rocha, R. 2011. An evidence-based model of distributed software development project management: results from a systematic mapping study. *Journal of Software Maintenance and Evolution: Research and Practice*, pp.n/a-n/a

Dieste, O. & Padua, O. a. G. Developing Search Strategies for Detecting Relevant Experiments for Systematic Reviews. *In: Empirical Software Engineering and Measurement*, 2007. ESEM 2007. First International Symposium on, 20-21 Sept. 2007 2007.

Glass, R. L., Vessey, I. & Ramesh, V. 2002. Research in software engineering: an analysis of the literature. *Information and Software Technology*, 44, pp.491-506

Gregor, S. 2006. The Nature of Theory in Information Systems. *MIS Quarterly*, 30, pp.611-642

Jalali, S. & Wohlin, C. Agile Practices in Global Software Engineering - A Systematic Map. In: Global Software Engineering (ICGSE), 2010 5th *IEEE International Conference* on, 23-26 Aug. 2010 2010.

Kitchenham, B. & Charters, S. 2007. Guidelines for performing Systematic Literature Reviews in Software Engineering.

Mohd Fauzi, S. S., Bannerman, P. L. & Staples, M. Software Configuration Management in Global Software Development: A systematic map. *In*, 2010.

Morrison, J. & George, J. F. 1995. Exploring the software engineering component in MIS research. *Commun. ACM*, 38, pp.80-91

Myers, M. D. & Klein, H. K. 2011. A set of principles for conducting critical research in information systems. *MIS Q.*, 35, pp.17-36

Orlikowski, W. J. & Baroudi, J. J. 1991. Studying Information Technology in Organizations: Research Approaches and Assumptions. *Information Systems Research*, 2, pp.1-28

Petersen, K., Feldt, R., Mujtaba, S. & Mattsson, M. Systematic mapping studies in software engineering. *In: Proceedings of the 12th international conference on Evaluation and Assessment in Software Engineering*, 2008 Italy. British Computer Society.

Portillo-Rodríguez, J., Vizcaíno, A., Piattini, M. & Beecham, S. 2012. Tools used in Global Software Engineering: A systematic mapping review. *Information and Software Technology*, 54, pp.663-685

Richardson, I., Casey, V., Mccaffery, F., Burton, J. & Beecham, S. 2012. A Process Framework for Global Software Engineering Teams. *Information and Software Technology*,

Šmite, D., Wohlin, C., Feldt, R. & Gorschek, T. Reporting empirical research in global software engineering: *A classification scheme*. In, 2008.

Šmite, D., Wohlin, C., Gorschek, T. & Feldt, R. 2010. Empirical evidence in global software engineering: a systematic review. *In Empirical Software Engineering*, 15, pp.91-118

Smith, M. A., Shneiderman, B., Milic-Frayling, N., Rodrigues, E. M., Barash, V., Dunne, C., Capone, T., Perer, A. & Gleave, E. 2009. Analyzing (social media) networks with NodeXL. *Proceedings of the fourth international conference on Communities and technologies*. University Park, PA, USA: ACM.

Steinmacher, I., Chaves, A. P. & Gerosa, M. A. 2012. Awareness Support in Distributed Software Development: A Systematic Review and Mapping of the Literature. *Computer Supported Cooperative Work: CSCW: An International Journal*, pp.1-46